# Development of fundamental power coupler for C-ADS superconducting elliptical cavities*

Kui-Xiang Gu (谷魁祥)[1, 2;1)]  Feng Bing (邝丰)[1, 2]  Wei-Min Pan (潘卫民)[1]  Tong-Ming Huang (黄彤明)[1]
Qiang Ma (马强)[1]  Fan-Bo Meng (孟繁博)[1]

[1] Institute of High Energy Physics, Chinese Academy of Sciences, Beijing 100049, China

[2] University of Chinese Academy of Sciences, Beijing 100049, China

**Abstract:** 5-cell elliptical cavities were selected for the main linac of China Accelerator Driven sub-critical System (C-ADS) in the medium energy section. According to the design, each cavity should be driven with radio frequency (RF) energy up to 150 kW by a fundamental power coupler (FPC). And as the cavity working with high quality factor and high accelerating gradient, the coupler should keep the cavity from contamination in assembly procedure. To fulfil the requirements, a single-window coaxial type coupler, featured in capability of handling high RF power, class 10 clean room assembly and heat load controlling, was designed. This paper presents how the coupler was designed and gives the details of RF design, heat load optimization and thermal analysis as well as multipacting simulations. In addition, a primary high power test is accomplished and described in this paper.

## 1   Introduction

The China Accelerator Driven sub-critical System (C-ADS) project is a CW proton linac consisting of two injector sections and one main linac. In the main linac, 650 MHz β=0.63 and β=0.82 5-cell elliptical superconducting cavities are applied to accelerate the beam current from the energy of about 180 MeV up to 1.5 GeV [1]. These cavities were designed to work at a high quality factor of $1\times10^{10}$ and a high accelerating gradient of 15 MV/m with a beam current of 10 mA [2]. To sufficiently drive each cavity, a fundamental power coupler (FPC) with the capability of delivering 150 kW RF power in CW was required.

The requirements of high quality factor and high working gradient of the cavities request that the coupler design should enable clean assembly procedures to minimize the risk of contaminating superconducting cavity (SC) [3]. One approach to satisfy this demand is double-window design such as TTF III couplers [4]. While referring to the present double-window couplers worldwide, the usage of cold window limits the power handling capability [5-7]. Another approach is a single-window design with the vacuum section of the coupler short enough to facilitate clean assembly, taking the design of SNS SC coupler for example [8]. However, the operation power of SNS coupler was 53 kW in average and the highest power record of this kind couplers was 125 kW, CW, made by BNL SRF Gun Coupler in test [9, 10]. And since coupler serves as a temperature bridge from the room temperature to the cryogenic temperature, short vacuum section and relatively high RF heating in proportion to RF power would result in excessive heat load. While for C-ADS where hundreds of FPCs are needed, very small heat loss is desired in the design to minimize operation costs [11]. The main challenge is to design a coupler which enable clean assembly, high power handling and low heat load simultaneously.

A coaxial type single-window FPC design was selected. Table 1 shows the main parameters of the cavities and the FPC. The design of the coupler was derived from the 500 MHz FPC for BEPCII SC cavity, on the account of its simplicity in structure and its capability of operating with high RF power [12]. The main dimensions were rescaled to 650 MHz; some components were redesigned based on the clean assembly procedures and cooling considerations. To minimize coupler's contribution to the overall heat load of the cryostat, the coupler outer conductor is copper plated,

* Supported by China ADS Project (XDA03020600)

1) Email: gukx@ihep.ac.cn

double-walled and cooled by helium gas. Thoroughly and carefully simulations are carried out to optimize the RF structures and heat load, decide cooling design and predict multipacting activities. A prototype was fabricated and a primary power test was accomplished. The details of simulations and the power conditioning is presented in this paper.

Table 1. Main parameters of the cavity and FPC

| Parameter | Value |
| --- | --- |
| Frequency/ MHz | 650 |
| Working gradient/ MV/ m | 15 |
| $Q_0$ (elliptical 082) | $1\times10^{10}$ |
| R/Q (elliptical 082)/ $\Omega$ | 514.6 |
| Beam current/ mA | 10 |
| Peak RF power/ kW | 150, CW mode |
| Coupling type | Antenna |
| Coaxial line Impedance/ $\Omega$ | 50 |

## 2  RF structure design

The FPC model is based on a 50 $\Omega$ coaxial line, including a coaxial planar window, a doorknob transition, an air side coaxial line and a vacuum side coaxial line, as shown in Fig. 1.

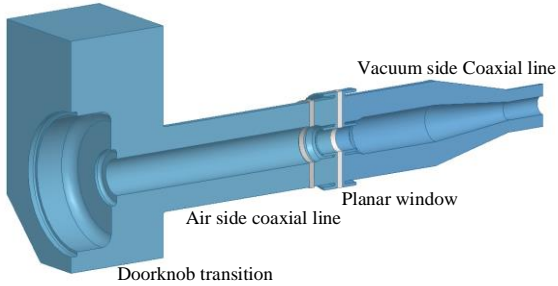

Fig. 1. FPC electromagnetic model

RF window is a critical component of the FPC. It works as a vacuum barrier for the cavity while let the RF power go through. The window is made of 97.6% $Al_2O_3$ content. To adequately handle high RF power, the window has a big diameter of 171 mm，brazed to a specially designed choke structure. The choke mainly contributes to impedance-matching of the window. Meanwhile, it weakens local electric field on the ceramic-copper weld joints which may be in dangerous when exposed to high intensity electric field, and presumably reduces the risk of multipacting. As shown in Fig. 2, the electrical field on the ceramic surface around the choke area is lower in contrast with the window with no choke. Specifically, on the surface of inner conductor to ceramic weld joint, the electrical field is 23.5 % lowered.

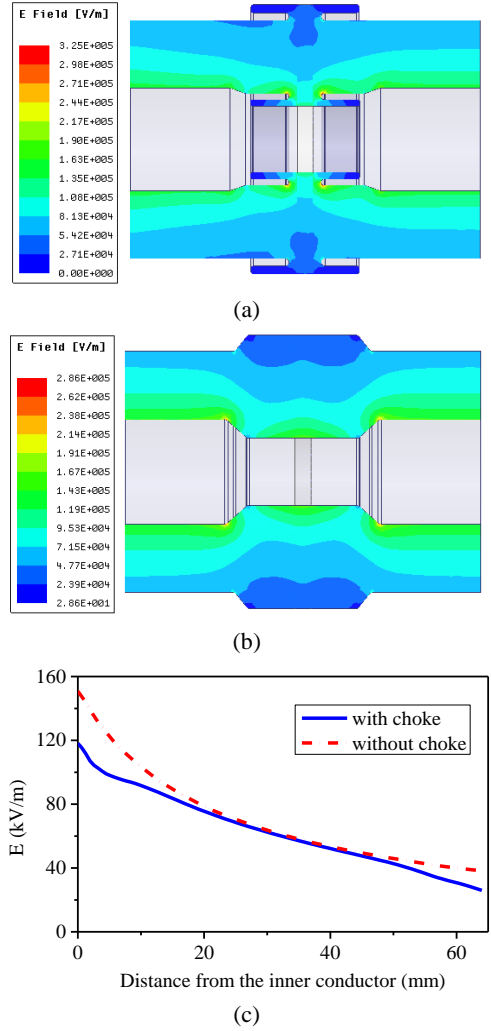

Fig. 2. The complex electric field contour of the window as 150 kW RF power go through: (a) window with choke structure; (b) window without choke structure. (c) The complex electric field on the surface of the ceramic in this two cases.

Transition from WR1500 waveguide which connecting to the klystron, to the coaxial line of the coupler is realized by a doorknob configuration. The RF performance of the doorknob is sensitive to its dimensions. In order to achieve a good impedance matching, dimensions of the doorknob were optimized, including the seat height (H), the seat diameter (D) and the distance from seat to the shot circuit

(L), as shown in Fig. 3.

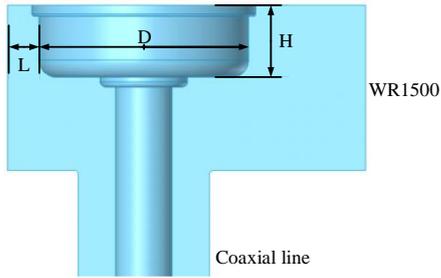

Fig. 3. The model of doorknob and its optimization dimensions.

Further simulations of the integrated coupler have been carried out. Fig. 4 shows the calculated S11 curve of the whole coupler. The S11 is about -46 dB at working frequency, 650 MHz. And the bandwidth is about 50 MHz at S11= -25 dB.

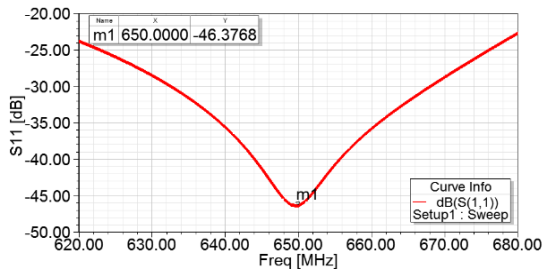

Fig. 4. Calculated S11 curve of the FPC ( S11= -46 dB at 650 MHz, relative permittivity of ceramic= 9.2).

## 3 Heat load optimization

As FPC perform as a thermal transition between room-temperature RF transmission system and the SC cryogenic environment, the coupler lead to static and dynamic heat load to the cavity and the cryogenic systems. For high power couplers, joule heat due to RF loss made the dynamic heat load considerable; meanwhile, the shorted vacuum section of the outer conductor (to enable clean assembly with cavity) greatly increased the static heat load. Therefore, how to reduce the coupler's contribution to the heat load to a reasonable low level becomes a big challenge.

In order to decrease RF losses, the stainless steel made outer conductor is copper plated. And the thickness of the copper plating which has a primary influence in the static heat load, was selected carefully. The copper plating should be thick enough to reduce RF dissipation while minimally increase thermal conduction as well as static heat load. Based on simulations and the copper plating technology, the thickness of the copper plating layer was chosen as 10 um, about three RF skin depths.

Then, cooling of the outer conductor was optimized carefully. Methods of cooling by thermo-anchors and helium gas were explored. In the case of thermo-anchor cooling, two thermo-anchors at 5 K and 80 K were attach to the outer conductor respectively. Analysis were accomplished to optimize the location of anchors for the sake of minimize total cryogenic load. The heat loads of the coupler are listed in Table 2. Obviously, the heat loads can't meet well with the cryogenic requirement. Therefore, the helium gas cooling was selected. As shown in Fig. 5, the outer conductor was a double-wall stainless steel tube with a spiral flow passage inside. The 5 K helium gas flow in the tube near the cold flange, spirel up and flow out close to the bottom of the RF window in 300 K. According to calculations, a flow rate of 0.05 g/s for helium gas is adequate. Correspondingly, the heat load of the FPC is less than 4 W (scaling to heat load at 5 K), smaller than the average heat loss of C-ADS CM1 FPCs [13].

Table 2. Heat flow at thermos-anchors for the coupler model with a 5 K and an 80 K anchor

| Heat load type | Heat load / (W) |
| --- | --- |
| 2 K heat load | 0.61 |
| 5 K heat load | 4.57 |
| 80 K heat load | 13.27 |

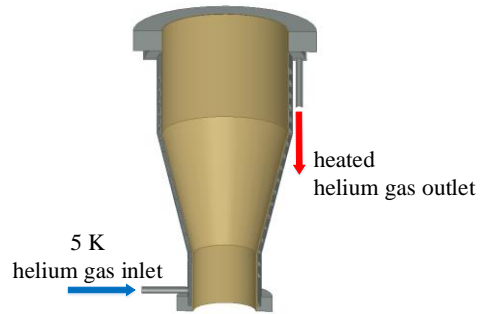

Fig. 5. The helium cooling outer conductor.

## 4 Thermal design

For FPCs work at high RF power, sufficient cooling are indispensable to control heat due to ohmic dissipation of

RF power in the walls of the coupler. Aim to secure that the coupler was adequately cooled, thermal analysis were performed for each component of the FPC.

The atmospheric components of the FPC including the doorknob and the air side coaxial line were both cooled by air flow, while the coaxial line was cooled in a special way. The outer conductor was machined with several rows of air outlets at the end nearing to the window. Therefore, the cold air which comes into doorknob can flow in the coaxial line and got both the inner conductor and outer conductor cooled.

The oxygen-free copper made inner conductor in the vacuum side was water cooled. Water entered through a tube which extend to the end of the inner conductor and flows along the internal wall of the inner conductor all the way up to the top of the window, taking away all the RF loss on the vacuum side inner conductor, a big fraction of dielectric loss in the ceramic and a part of RF loss on the air side inner conductor. To keep the cooling water temperature rise less than 1°C (entering at 25°C) with 150 kW RF power passing through, a flow rate of 2.5 l/min is required. As Fig. 6 shown, with air and water cooling, the highest temperature appears in the middle of the air side inner conductor or in the center of the ceramic. And the highest temperature does not exceed 37°C.

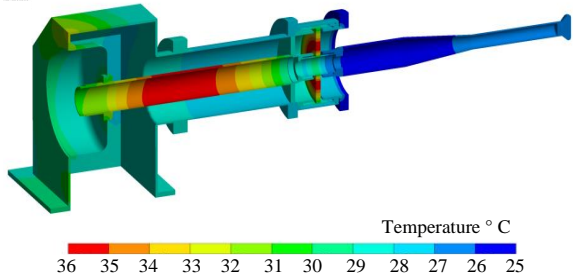

Fig. 6. Temperature contour of the FPC's air side conponents with 150 kW RF power passing through in traveling wave mode.

RF Window is the most fragile component of coupler as it is prone to high stresses due to thermal expansion and contraction. Thermal analysis and mechanical stress analysis were performed for window under 300 kW (twice the operational RF power). As Fig. 7 shows, the maximum temperature is 49.6°C, located in the center of the ceramic and the maximum stress is 56.9 MPa, appearing on the edges of ceramic-copper weld joints. The maximum stress is less than one fifth of the flexural strength of the ceramic.

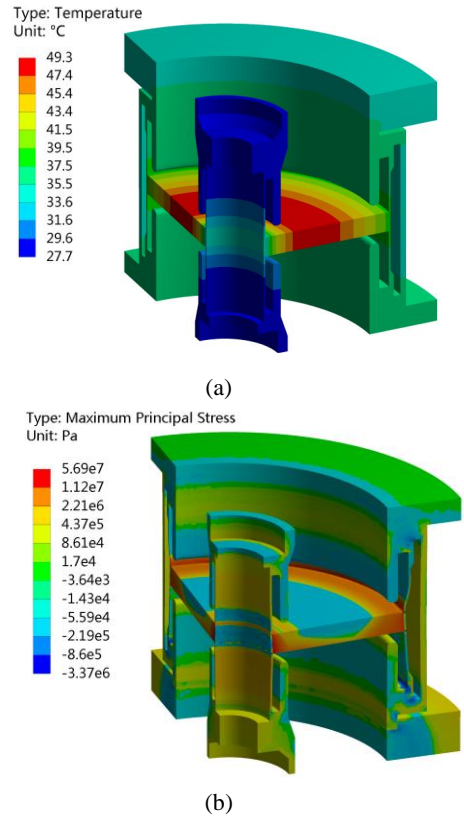

Fig. 7. (a) Temperature contour of the window assembly with 300 kW RF power through. (b) The corresponding maximum principle stress contour of the window assembly.

## 5  Multipacting simulations

Simulations of multipacting for the FPC were performed with MultiPac 2.1, an electron trajectory tracking code with a 2D FEM field solver [14]. The analysis was carried out for the vacuum side coaxial line and window structure respectively.

In the simulations, the RF power was swept from 1 kW to 150 kW with full reflection at different reflection phases and different emitting phases of the initial electrons. The simulation results of the coaxial line, as Fig. 8 shown, revealed that nearly all of the free electrons were disappeared after 20 times impacts and the final residual electrons were less than initial electrons at each power level below 150 kW. Simulation results of the window at different reflection phases turned out to be similar with that of the coaxial line. These simulations indicate that there is no hard multipacting barrier under 150 kW power for the FPC.

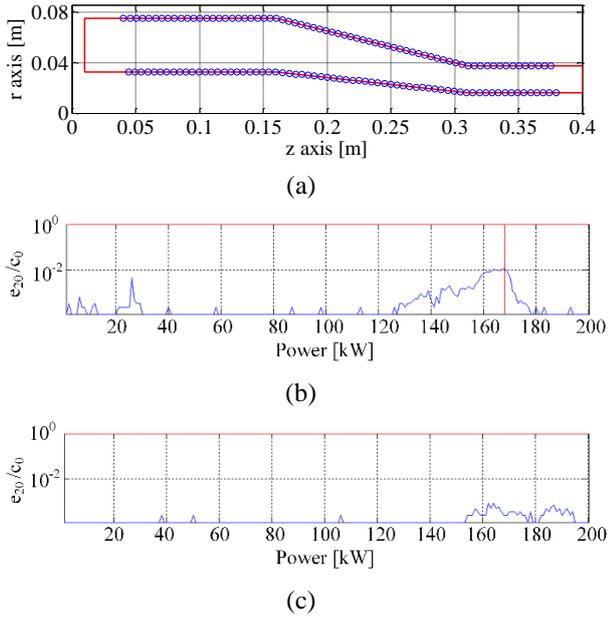

Fig. 8. (a) 2D model of the coaxial line with a taper for multipacting simulation in MultiPac 2.1. And the circles represent the initial electrons. (b) The enhanced electron counter function in the coaxial line with an electric wall at the right end of the model. (c) The enhanced electron counter function in the coaxial line with a magnetic wall at the right end of the model. Note that multipacting occurs only when the enhanced counter function is greater than one.

## 6  High power conditioning

One FPC was successfully manufactured and the FPC was tested with a 500 MHz klystron as the 650 MHz power source was not ready to work. To get 500 MHz RF power going through the FPC with little reflection, the original doorknob was replaced with a doorknob of BEPCII coupler [12]. In the test stand, The FPC was mounted on a connecting waveguide with an already conditioned BEPCII coupler. RP power came through the FPC in the upstream, went to the BEPCII coupler through the connecting waveguide and finally, was absorbed by a terminating load which connected to the downstream coupler. The conditioning stand was shown in Fig. 9.

The conditioning was started after a high temperature baking of the test stand for about 51 hours. Pulsed and CW conditioning method were applied in turns. During the RF processing, ARC signal, electron signal, vacuum and temperature of the couplers were monitored and interlocked with the power to prevent damage of the couplers. As was shown in Fig. 10, after five days of conditioning, the target power of traveling wave 150 kW in CW was reached. And there was no unexpected temperature rise observed on the FPC components in the test. The highest temperature on the window was recorded as 32.3°C, after kept on 150 kW conditioning for more than thirty minutes.

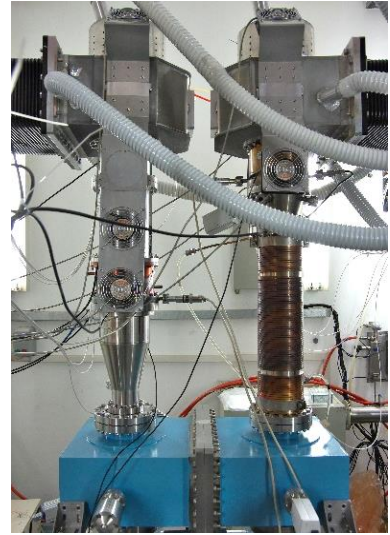

Fig. 9. Test stand used in condition. The left coupler was the 650 MHz C-ADS FPC with the doorknob replaced and the right one was a BEPCII FPC. The two couplers were connected through a waveguide cavity.

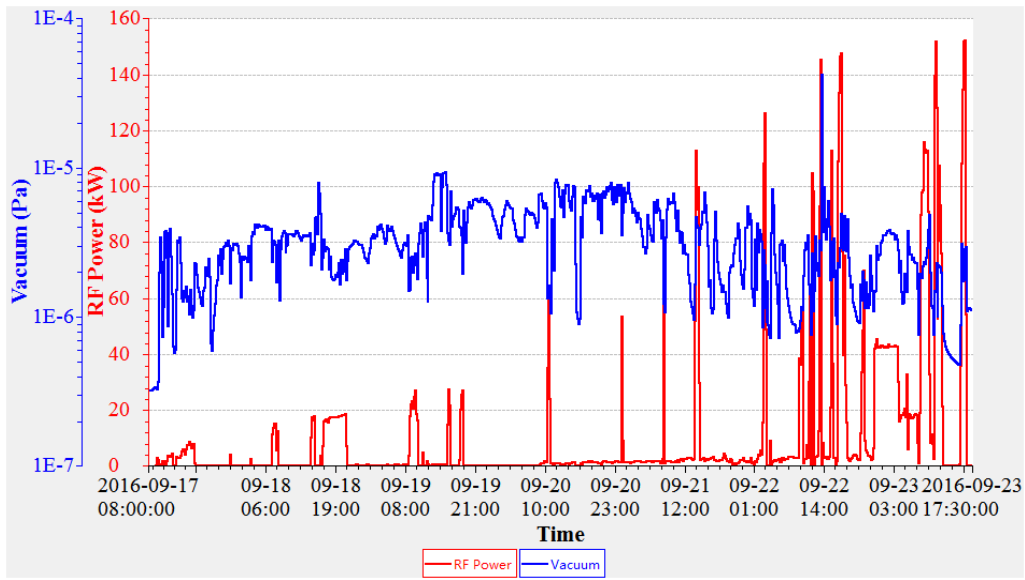

Fig. 10. Conditioning history of the FPC. Pulsed and CW conditioning method were applied in turns in the processing. RF power of 150 kW was reached after conditioning for about 5 days.

## 7   Summary

A new high power coupler for 650 MHz C-ADS superconducting elliptical cavities has been designed and tested. The FPC was design with a big-diameter planar window with choke structure to provide 150 kW of CW RF power to the cavities. The vacuum section is shorted to enable clean mounting with cavity and accordingly, copper plating thickness optimization and helium gas cooling are applied to minimize the heat load. Sufficient water and air cooling methods are adopted to cooling down the FPC. Multipacting simulations indicate that there is no hard multipacting barrier in the coupler. Moreover, a primary power test was finished and testified the FPC's power handling capability. We believe that this work will provide a reference for future design of similar high power couplers which serve to superconducting cavities of high quality factor and high accelerating gradient.